# A model of quantum reality

Eduardo V. Flores[1]

**Here we propose a model of particles and fields based on the mathematical framework of quantum physics. Our model is an interpretation of quantum physics that treats particles and fields as physically real. We analyze four experiments on wave-particle duality that pose paradoxes. We show that within our model these paradoxes are resolved.**

Is quantum reality out of this world? A reading of a recent paper by XS Ma *et al.*[1] could convince us that this is indeed the case. In their experiment XS Ma *et al.* show how a photon that has already been detected could be chosen to exhibit either particle or wave properties. The choice of what to observe is made by a distant observer who is so far away as not to influence the outcome for the already detected photon. Their results appear so convincing that the authors conclude:

*No naïve realistic picture is compatible with our results because whether a quantum could be seen showing particle- or wave-like behavior would depend on a causally disconnected choice. It is therefore suggestive to abandon such pictures altogether.*[1]

In fact, this is just one of many experiments on the foundations of quantum physics that question our naïve notion of reality[2-5]. While experimental results[6] such as superluminal propagation of correlations shown in the violation of Bell's inequalities are surprising others that deal with the nature of an object are paradoxical[1,3,7]. How could our causally disconnected choice make an already detected object be either wave or particle? Here we present a model that appears to solve wave-particle duality paradoxes and is based on the mathematical formulation of quantum physics.

**The model**
In our model we adopt the mathematical formulation of quantum field theory and in particular quantum optics. We make assumptions about particles and fields. Our choice of assumptions is justified by their suitability in resolving paradoxes. Our assumptions are:

  I.   The particle is a real point-like object at all times
 II.   The quantum field is physically real
III.   The quantum field is not the particle
 IV.   The particle is driven by the quantum field
  V.   The location of the particle within the quantum field is probabilistic in such a way that where the field is stronger the particle is more likely to be found

Our model has similarities with Bohmian mechanics. Both models are quantum theories describing the motion of point-like particles with definite trajectory[8,9]. In our model the actual location of the particle does not affect the development of the quantum field. Similarly, in Bohmian mechanics the location of the particle does not affect the development of the wavefunction. However, there are major differences. In our model the conservation laws help determine particle trajectory but we do not attempt to give an equation of motion for the actual trajectory. In Bohmian mechanics the position changes according to an equation of motion known as the guiding equation. In our model we deal with the quantum field. On the other hand, in Bohmian mechanics the wavefunction is central. Our model starts at the relativistic level while Bohmian mechanics has been successful at non-relativistic level.

---

[1] Department of Physics & Astronomy, Rowan University, Glassboro, NJ 08028. e-mail: flores@rowan.edu

We notice that in our model a particle has definite position and momentum; this appears to be in contradiction to Heisenberg's uncertainty principle. However, Heisenberg's uncertainty principle deals with the limits of accuracy in a measurement of conjugate variables such as momentum and position. Just because we cannot measure the exact momentum and position of a particle it does not imply that there is no exact momentum and position for that particle[10]. Similarly, we observe that in our model we have a real particle and a real wave a fact that appears to be in contradiction to the complementarity principle. However, the complementarity principle deals with limits in a measurement of wave and particle aspects. The analysis of the quantum erasure experiment below will show how it is possible to have a real particle and a real wave and yet not be able to measure both properties simultaneously. Our model contradicts neither the uncertainty principle nor the principle of complementarity when both refer to actual measurements.

In a recent article Pusey *et al.* present a no-go theorem[11], which shows that models in which the quantum state corresponds to mere information about a physical state cannot reproduce the predictions of quantum theory. In our model the quantum state of the system is the quantum field. In our work we assume that the quantum field is real, thus, it corresponds directly to reality and our model passes Pusey *et al.* test. However, Pusey *et al.* also point out a problem with a physical quantum state that collapses on measurement. Fortunately, the quantum field does not necessarily collapse on measurement, as it will be shown below in the analysis of the particle and the two slits experiment.

**Real empty wave packets**
We first analyze the delayed choice experiment originally proposed by Wheeler[2]. A highly attenuated beam enters a Mach-Zehnder setup as in Fig. 1 and reaches a 50:50 beam splitter. We assume we have a single photon. Right before the photon reaches the intersection, where the beams cross, we have the choice of placing or not placing a second beam splitter. In setup 1 we do not place the second beam splitter and the photon ends up at one of the detectors as in Fig. 1. Applying energy-momentum conservation we extrapolate the path of the photon from detector all the way to the initial entrance and obtain full path information.

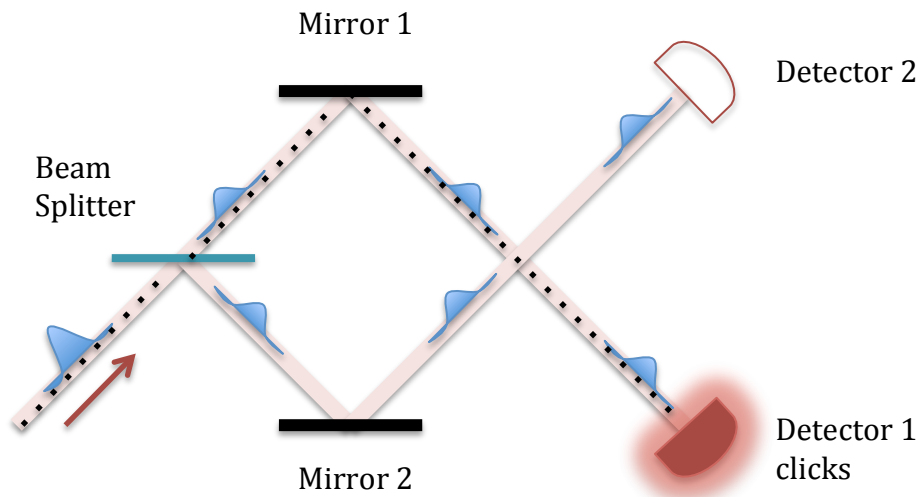

**Fig. 1| Set up 1 for the delayed choice experiment.** If detector 1 clicks we obtain particle information by tracing the path of the particle applying energy-momentum conservation. In this figure, the particle takes the path shown by the dots. In our model, at the fixed beam splitter, the incoming quantum field wave packet splits into two smaller wave packets; one is reflected and takes one path and the other is transmitted and takes the alternative path. The empty wave packet that reaches detector 2 does not activate it.

In setup 2 a second beam splitter is inserted as in Fig. 2. Setup 2 is arranged so that destructive interference is produced along the path that reaches one detector and constructive interference is produced along the other path. In this case the photon will end up at the detector with the path that shows constructive interference. In the delayed-choice experiment we wait until the last picosecond before the photon reaches the intersection to make the choice of whether to use setup 1 or setup 2. We note that the delayed choice experiment has been experimentally realized[7].

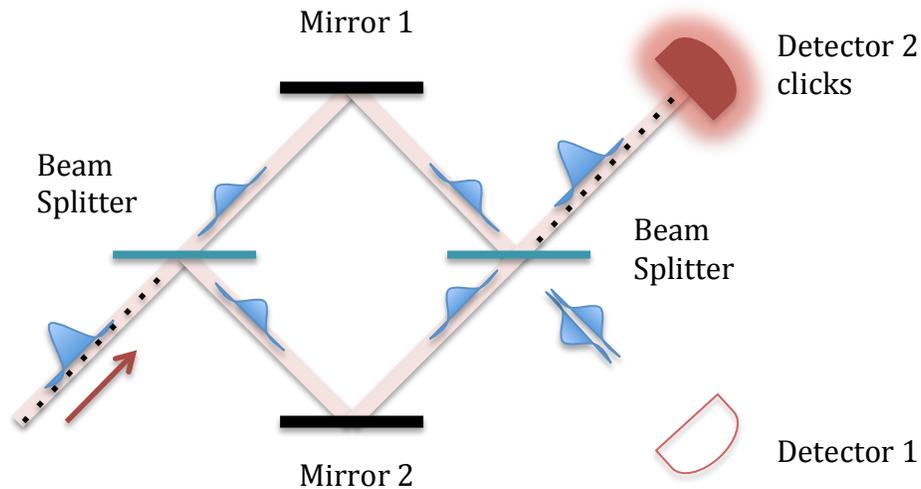

**Fig. 2| Set up 2 for the delayed choice experiment.** The second beam splitter produces destructive interference along the path that ends at detector 1. The particle is driven to the constructive interference region, which is the path that ends at detector 2. We cannot tell which of the two paths within the beam splitters the particle actually takes; Wheeler assumes that both paths are taken. However, in our model the particle takes a single path, not shown in this figure.

According to Wheeler, the consequences of our choice are paradoxical. If we choose setup 1 the photon has taken a single path. If we choose setup 2 the photon has taken both paths. What is striking here is that choice of the observer affects the past history of the photon:

*There (delayed-choice experiment) we make the decision whether to put the final half-silver mirror in place or to take it out at the very last picosecond, after the photon has already accomplished its travel. In this sense, we have strange inversion of the normal order of time. We, now, by moving the mirror in or out have an unavoidable effect on what we have a right to say about the already past history of that photon.*[12]

Here is how we understand this experiment within our model. The particle and the accompanying quantum field wave packet enter the Mach-Zehnder setup as in Fig. 1. The 50:50 beam-splitter splits the wave packet equally while the point-like particle takes one of the two equally likely paths. Later the beams cross each other. Right before the photon reaches the intersection we have the choice of placing or not placing a second beam splitter. If we do not place the second beam splitter the particle would activate a detector. Applying energy-momentum conservation we extrapolate the path of the particle from excited detector all the way to the initial entrance and obtain full path information. We note that in our model an empty wave packet cannot activate a detector. Had we placed the second beam splitter as in Fig. 2 we would never see the particle reaching one of the detectors due to destructive interference. We note that in this case, within the beam splitters, the actual path of the particle is unknown; however, in our model the particle takes a single path.

Therefore, according to our model there is no paradox. Moving in or out the second beam-splitter has no effect on the past history of that photon. In setup 1 and setup 2 the wave packets always occupy the two available paths within the beam splitters while the particle is always in a single path.

**The wave is not the particle**

In 1967 Pfleegor and Mandel reported[3] that they observed interference of single photons from independent photon beams. The Pfleegor-Mandel experiment consists of the two independent lasers with beams aligned so that they intersect at a small angle and interfere as in Fig. 3. However, the interference patterned they observed shifted from trial to trial. Using a different setup Radloff[13] obtained a stable interference pattern that could be built at a rate low enough to ensure the presence of a single photon at a time.

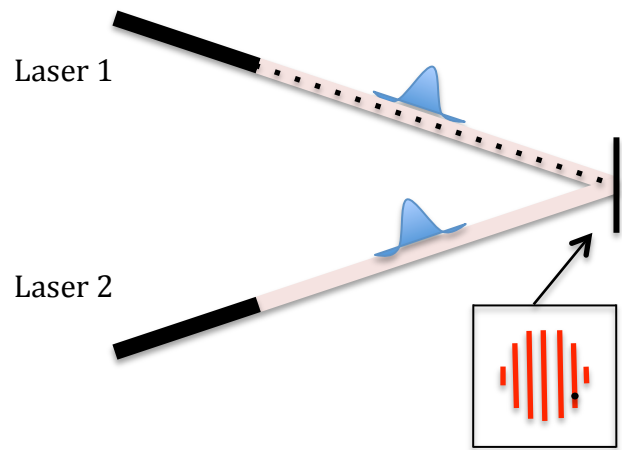

**Figure 3| Independent photon beams experiment.** Two independent photon sources but only one photon is present in the entire setup. At the intersection an interference pattern is formed one particle at a time. In our model one beam has an empty wave packet that interferes with the wave packet from the other beam. In this figure the particle comes from laser 1 and eventually contributes as a dot to the interference pattern.

Pfleegor and Mandel associate the formation of the interference pattern with the intrinsic uncertainty from which of the two sources the photon comes from[3]. However, Pfleegor and Mandel interpretation implies that the quantum state represents mere information, which is a paradoxical results in the light of Pusey *et al.* no-go theorem mentioned earlier.

H. Paul in an extensive review of the subject presents an alternative interpretation:

*The proper description of interference between independent photons will be as follows. What interferes with one another are waves, ... What actually happens in the detection process is that an energy packet hv is taken from the superposition field to which both lasers contribute equally.*[14]

We note that interference of the wave from a photon in one beam with the wave from a photon in the other beam runs into conflict with Dirac's dictum:

*... each photon interferes only with itself. Interference between different photons never occurs.*[15]

We may claim that Dirac's dictum is based on limited experimental evidence of his time. Nevertheless, we can find other reasons for this dictum to be proposed. According to particle-wave duality, particle and wave are equally valid complementary aspects of a photon. Just as the particle can be identity uniquely as a single dot on a detector, so too, the wave could be identity uniquely by interfering only with itself. Thus, the fact that waves from two independent photons interfere remains paradoxical.

The second part of H. Paul's statement on interference of independent photons deals with the source of the energy packet $h\nu$. In H. Paul's view the photon is cogenerated by two independent sources. Half of the photon is taken from one beam and the other half is taken from the other. The fact that independent beams depend on each other for the formation of the observed photon means that they are not independent. This is another paradox.

Here is how we resolve the present paradoxes. In our model, a wave packet is not the particle; thus, the origin or identity of the wave is unimportant. A particle could respond to any relevant wave present. Since the wave is not identified with the particle we are not bound by Dirac's dictum. Therefore, we predict that any two optical fields independent or not would interfere; this is actually an experimental fact[14,16].

Now we consider the situation when we only have one photon in the entire setup. At this level there may be a single particle but there would be empty quantum field wave packets from both sources that continue to propagate. We note that the presence of empty wave packets was encountered before in the delayed choice experiment. There, at the first beam splitter, while the particle takes one of the two paths, the wave packet splits into two wave packets. Thus, in one of the paths there is an empty wave packet. In the experiment with independent photon beams wave packets from both sources meet at the intersection and interfere as in Fig. 3. The particle that comes from either source would most likely show up at a place where constructive interference takes place.

**Complementarity is not violated**

In the quantum erasure with causally disconnected choice experiment two entangle photons are produced, one we call system and the other environment[1]. We prepare the polarized entangled state

$$|\chi\rangle = \frac{1}{\sqrt{2}}(|H\rangle_s|V\rangle_e + |V\rangle_s|H\rangle_e) \tag{1}$$

where $|H\rangle$ and $|V\rangle$ denote quantum states of horizontal and vertical linear polarization, and *s* and *e* index the system and environment photon, respectively. We may change this state to the R/L basis by replacing $|H\rangle = (|R\rangle + |L\rangle)/\sqrt{2}$ and $|V\rangle = i(|L\rangle - |R\rangle)/\sqrt{2}$. Thus, we have two alternative choices for how to represent the state in Eq. 1

$$|\chi\rangle = \begin{cases} \frac{1}{\sqrt{2}}(|H\rangle_s|V\rangle_e + |V\rangle_s|H\rangle_e), & \text{choice 1} \\ \frac{i}{\sqrt{2}}(|L\rangle_s|L\rangle_e - |R\rangle_s|R\rangle_e), & \text{choice 2} \end{cases} \tag{2}$$

The system photon is sent through a polarizing beam splitter that transmits the horizontal and reflects the vertical polarization state. The vertical polarization state is sent along path *a* and the horizontal polarization state is sent along path *b*. Two in-line polarization controllers rotate the orthogonal polarization states of the photon in path *a* and *b* to an identical one thus eliminate the polarization distinguishability of the two paths. Therefore, we replace $(|H\rangle_s \rightarrow |b\rangle_s)$ and $(|V\rangle_s \rightarrow |a\rangle_s)$ in Eq. 2 and obtain

$$|\chi\rangle = \begin{cases} \dfrac{1}{\sqrt{2}}(|b\rangle_s|V\rangle_e + |a\rangle_s|H\rangle_e), & \text{choice 1} \\ \dfrac{1}{2}[(|a\rangle_s + i|b\rangle_s)|L\rangle_e + (|a\rangle_s - i|b\rangle_s)|R\rangle_e], & \text{choice 2} \end{cases} \quad (3)$$

Paths $a$ and $b$ eventually recombine at a beam splitter and the system photon is detected at either of two detectors placed in front of each incoming beam. On the average each detector gets the same number of photons. The fascinating aspect of this experiment is that we are able to sort out two complementary interference patterns out of this random collection of system photons. The information needed to sort out the interference patterns depends on measurements on the environment photon and the applicability of Eq. 3. We measure the polarization state of the environment photon under the condition that this measurement and the choice of what to measure are causally disconnected from the events that happen to the system photon. We note that the applicability of Eq. 3 under these conditions implies superluminal signalling, of the kind seen in the violation of Bell's inequalities[6].

If, in choice 1, we find that the environment photon polarization was horizontal $|H\rangle_e$ then we would know that according to Eq. 3 the system photon state would be $|a\rangle_s$, thus, it would take path $a$. Similarly, if the environment photon was found with polarization $|V\rangle_e$ then we would know that the system photon state would be $|b\rangle_s$, thus, it would take path $b$. Since, in choice 1, we identify the actual path of the system photon it does not form part of an interference pattern. We note that in choice 1 we identify the photon path but we erase wave information since we would not be able to tell to which of the two interference patterns the system photon belongs.

If, in choice 2, we find that the environment photon circular polarization was left $|L\rangle_e$ then we would know that according to Eq. 3 the system photon state would be $(|a\rangle_s + i|b\rangle_s)/\sqrt{2}$; thus, the system photon would take both paths. This photon would be part of an interference pattern with fringes. Similarly, if the environment photon was found with polarization $|R\rangle_e$ then we would know that the system photon state would be $(|a\rangle_s - i|b\rangle_s)/\sqrt{2}$; thus, this photon too would take both paths and would form part of an interference pattern with anti-fringes. We note that in choice 2 we obtain wave information but we erase path information since we would not be able to tell to which of the two paths the system photon would take.

In summary, once we have measured the system photon and kept its record, we could, in a distant future at a place causally disconnected from the system photon, decide what had happened to that system photon by simply measuring the state of polarization of the environment photon. If we choose to observe the linear polarization of the environment photon we would find that the system photon took a single path either $a$ or $b$. If we choose to observe the circular polarization of the environment photon we would find that the system photon took both paths $a$ and $b$. This is a paradox.

Once again, the paradox is solved within our model once we accept the simultaneous reality of wave and particle. The system photon enters a polarizing beam splitter that leads to paths that eventually recombine. In every trial the system particle takes a single path while the accompanying quantum field wave packet splits and takes both paths. The two possible paths for the system particle are $a$ and $b$. The two possible phases for the accompanying wave packet are $|a\rangle_s + i|b\rangle_s$ and $|a\rangle_s - i|b\rangle_s$. Thus, there are four possible cases for the system photon shown in Fig. 4.

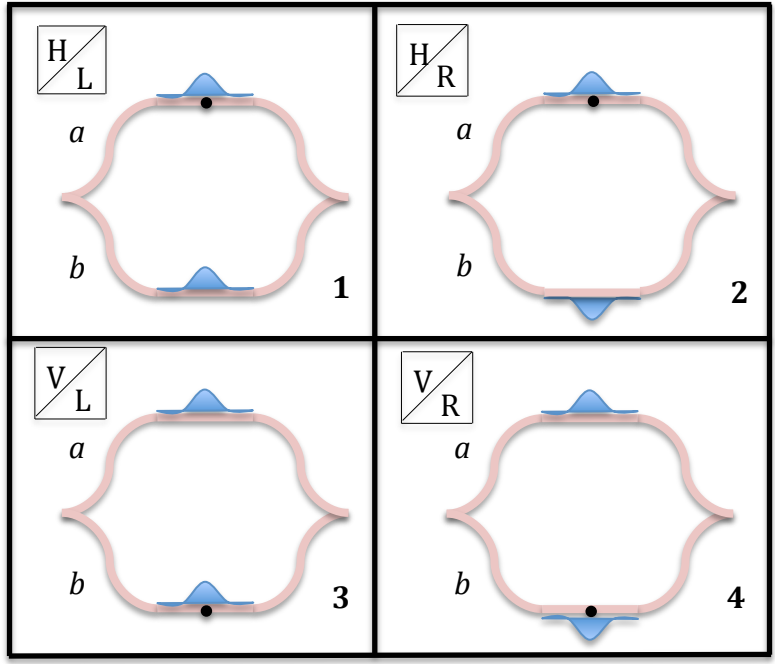

**Figure 4| Four possible cases for the system photon.** In each case the system photon enters from the left to a polarizing beam splitter. The accompanying wave packet is equally split with two possible phases. The particle, described by the dot, takes either path $a$ or $b$. The paths eventually recombine. The upper left corner box in each case shows what the environment photon would show, according to Eq. 3, upon measurement of its linear ($H$ or $V$) or circular ($L$ or $R$) polarization.

We see in Fig. 4 that if the environment photon was found in state $|L\rangle_e$ then the indistinguishable cases would be 1 and 3; thus, the path of the particle would be unknown. According to Eq. 3, the corresponding system photon would contribute to an interference pattern with fringes. If the environment photon was found in state $|R\rangle_e$ then the indistinguishable cases would be 2 and 4; thus, the path of the particle would be unknown. According to Eq. 3, the corresponding system photon would contribute to an interference pattern with anti-fringes. If, on the other hand, the environment photon was found in state $|H\rangle_e$ then we would know that the system photon took path $a$ and the indistinguishable cases would be 1 and 2; thus, we could not distinguish to which interference pattern the corresponding system photon belongs. Similarly, if the environment photon was found in state $|V\rangle_e$ then we would know that the system photon took path $b$ and the indistinguishable cases would be 3 and 4; thus, we could not distinguish to which interference pattern the corresponding system photon belongs. Therefore, what the environment photon reveals upon measurement is not paradoxical but what is expected when particle and wave are simultaneously real and Eq. 3 is obeyed. We learn from this this experiment that complementarity is not violated even when particle and wave are simultaneously real.

**The quantum field does not collapse on measurement**
We apply our model to the particle and two slits experiment. Let us consider a very large absorbent wall with two pinholes. We assume that the source of the particle is at infinity so that any quantum field wave packet accompanying the particle has been dissolved. Thus, we are working at the first excited state of the vacuum field represented by a plane wave. The quantum vacuum has non-zero field modes of all frequencies and directions. One of the components of this field is a wave with wavenumber $k$ moving to the right. We know that vacuum field modes are shaped by boundary conditions[17]. The vacuum field mode in Fig. 5 interacts with the wall with pinholes. Left of the wall the field is a plane wave. Right of the wall the field has two cones emanating out of the pinholes. The cones will intersect and form regions of constructive and destructive interference. We note that if one of the pinholes was closed then we would only observe a single cone and no interference.

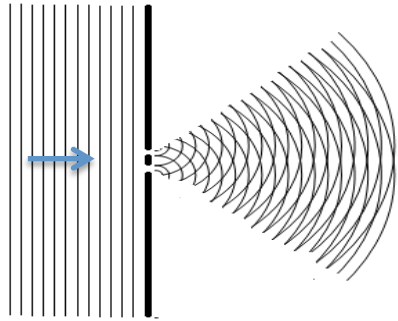

**Fig. 5| A vacuum field mode shaped by boundary conditions.** An absorbent wall with two pinholes shapes a field mode represented by a plane wave propagating to the right. Left of the wall we have a plane wave and right of the wall there are regions of constructive and destructive interference.

Now, we assume that the particle with momentum that corresponds to wavenumber $k$ approaches the wall from the left. The incoming particle occupies the field mode depicted in Fig. 5. The particle might go through one of the two pinholes and eventually hit a detector screen located right of the wall. Along its path, the particle is driven by the field mode; it is kept away from regions of destructive interference and it is lead to regions of constructive interference. Thus, it is more likely that the particle would hit the screen at a place where constructive interference occurs. The hit would be recorded as a point on the screen.

When the particle hits the screen it leaves a mark, which constitutes a measurement. At measurement, the particle vacates the field mode described in Fig. 5 and could occupy a new field mode or could get absorbed. The particle could end up in a bound state mode due to the elements that form the screen. We note that the field mode in Fig. 5 would be unchanged after being vacated by the particle. This quantum field does not collapse on measurement. If another particle with identical momentum to the first approaches the wall it would occupy the mode described in Fig. 5. The particle could be detected as another point on the screen. The collection of many identically prepared particles would eventually form on the screen an interference pattern with bright and dark fringes.

If we decide to investigate through which pinhole the particle goes through we could block one of the pinholes. When a pinhole is blocked we have changed boundary conditions, which results in a different field mode. Now, right of the wall, there would be a single cone that by itself produces no interference. However, we may decide to obtain information about which pinhole the particle crosses using a mechanism that partially blocks the pinholes. Imagine that the particles that come from far away were electrons and a source of photons was placed behind the pinholes. The setup would be such that whenever a scattered photon would be detected we would know which pinhole the electron crossed[4]. As it is well known this does not work either. The uncertainty principle keeps us from getting information about which pinhole each electron crosses through while simultaneously maintaining an interference pattern.

Richard Feynman questions the existence of a mechanism that would explain the odd results of the particle and the two slits experiment[4]. According to classical physics, the probability density obtained with two slits open is the sum of probability densities obtained by opening one slit at a time. In quantum mechanics it is different; the probability density obtained with two slits open is not the sum of probability densities obtained by opening one slit at a time. This is Feynman's only mystery of quantum mechanics[4]. However, this mystery may have a solution. We note that here we have two different boundary conditions: one pinhole open and two pinholes open. Two different boundary conditions generate two different field configurations one with an interference pattern and one without. When both slits are open the interference pattern is already there even if there were no electrons to show it. Since the quantum field drives the particle we expect that the actual experiment with electrons would show the interference pattern when both slits are open and no interference pattern when a single slit is open. Thus, it appears to us that appealing to the physical reality of particle and wave solves Feynman's mystery of quantum mechanics.

**Prospects**

We have proposed a model of quantum physics based on the mathematical framework of the theory. In our model wave and particle are physically real. The particle is a point-like object. The wave is the quantum field, which has a dominant role on the dynamics of the particle. We have analyzed four experiments that pose particle-wave duality paradoxes and show how our model appears to resolve the paradoxes. Our main concern with long standing unresolved paradoxes is their power to hinder our ability to make progress in the field. We invite others to examine our model and test it on relevant experiments. An important aspect of our model is that it appears to pass Pusey *et al.* no-go theorem as our model is based on a real quantum field that does not collapse upon measurement. However, there is work to be done such as working out the details of the non-relativistic version of our model that could lead to a clearer interpretation of the wavefunction when applicable. We would like to know to what extend particle trajectory could be obtained from the conservation laws. As long-term projects, we would like to investigate how gravity fits our model, and the nature of the physical reality of the quantum field. More important: is there an experimental test for our model?